\documentclass[english]{revtex4-1}
\usepackage{helvet}
\let\origrmdefault\rmdefault
\usepackage[math]{iwona}
\renewcommand{\rmdefault}{\origrmdefault}

\usepackage[T1]{fontenc}
\usepackage[latin9]{inputenc}
\usepackage{color}
\usepackage{bm}
\usepackage{amsmath}
\usepackage{graphicx}
\usepackage{setspace}

\makeatletter

\providecommand{\tabularnewline}{\\}

\usepackage{natbib}
\setcitestyle{super}
\usepackage{bm}

\newcommand{\uv}{\mathbf{u}}
\newcommand{\fv}{\mathbf{u}}

\newcommand{\qobar}{\bar{q}}

\newcommand{\cbar}{\bar{c}}

\makeatother

\usepackage{babel}
\begin{document}
\title{{\Large{}Normalization of the task-dependent detective quantum efficiency
of spectroscopic x-ray imaging detectors}}
\author{Jesse Tanguay$^{a}$ and Mats Persson$^{b,c}$}
\affiliation{a) Department of Physics, Ryerson University, Toronto, CA, M5B 2K3}
\affiliation{b) Department of Physics, KTH Royal Institute of Technology, 114 28
Stockholm, Sweden}
\affiliation{c) MedTechLabs, BioClinicum, Karolinska University Hospital, 171 64
Solna, Sweden}
\begin{abstract}
\noindent \textbf{Purpose: }Spectroscopic x-ray detectors (SXDs) are
poised to play a substantial role in the next generation of medical
x-ray imaging. Evaluating the performance of SXDs in terms of the
detective quantum efficiency (DQE) requires normalization of the frequency-dependent
signal-to-noise ratio (SNR) by that of an ideal SXD. The purpose of
this article is to provide mathematical expressions of the SNR of
ideal SXDs for quantification and detection tasks and then to tabulate
their numeric values for standardized tasks and standardized x-ray
spectra.\\
\textbf{Methods: }We propose using standardized RQA x-ray spectra
to evaluate the performance of SXDs. We define ideal SXDs as those
that (1) have an infinite number of infinitesimal energy bins, (2)
do not distort the incident distribution of x-ray photons in the spatial
or energy domains, and (3) do not decrease the frequency-dependent
SNR of the incident distribution of x-ray quanta. We derive analytic
expressions for the noise power spectrum (NPS) of such ideal detectors
for detection and quantification tasks.\textbf{ }We tabulate the NPS
of ideal SXDs for RQA x-ray spectra for detection and quantification
of aluminum, PMMA , iodine, and gadolinium basis materials.\textbf{}\\
\textbf{Results: }Our analysis reveals that a single matrix determines
the noise power of ideal SXDs in detection and quantification tasks,
including basis material decomposition and line-integral estimation
for pseudo-mono-energetic imaging. This NPS matrix is determined by
the x-ray spectrum incident on the detector and the mass-attenuation
coefficients of the set of basis materials (e.g.\textbf{ }PMMA, aluminum
and iodine) to be detected or quantified. For a set of $N$ basis
materials, the NPS matrix of the ideal SXD is an $N\times N$ symmetric
matrix.\textbf{ }Combining existing tabulated values of the mass-attenuation
coefficients of basis materials with standardized RQA x-ray spectra
enabled tabulating numeric values of the NPS matrix for selected spectra
and tasks.\textbf{}\\
\textbf{Conclusions:} The numeric values and mathematical expressions
of the NPS of ideal SXDs reported here can be used to normalize measurements
of the frequency-dependent SNR of SXDs for experimental study of the
task-dependent DQE.
\end{abstract}
\maketitle

\section{Introduction}

Spectroscopic x-ray imaging detectors (SXDs) perform crude estimates
of the shape of diagnostic x-ray spectra,\citealp{Lundqvist2001,Chmeissani2004,Shikhaliev2006_2}
enabling single-shot basis-material decomposition, pseudo monoenergetic
imaging, and optimal energy weighting.\citealp{Roessl2007_2,Schlomka2008,Feuerlein2008,Bornefalk2011,Yveborg2015,Tao2019,Richtsmeier,Dunning2020,Danielsson2021}
An active research area is the development of frameworks for evaluating
the performance of SXDs.\citealp{Persson2018,Rajbhandary2018,Persson2019,Rajbhandary2020,Persson2020,Tanguay2021}
The challenge is two-fold: (1) SXDs record multiple images for each
exposure and these images are correlated with each other, and (2)
the spectral data recorded by SXDs can be used for a number of different
tasks, for example energy weighting, basis-material decomposition,
or pseudo-monoenergetic imaging.

Persson\emph{~et~al.}\citealp{Persson2018,Persson2019,Persson2020}
defined a detective quantum efficiency (DQE) for SXDs for detection
and quantification tasks. For detection tasks, the DQE quantifies
the ability of an ideal observer, i.e. one that has full knowledge
of the signal and noise properties of the imaging system, to use the
data provided by an SXD to detect a known signal in a known background.
This DQE is inherently task-dependent and can be calculated for a
signal difference resulting from any element of a set of basis materials,
for example water, bone and iodine. For quantification tasks, the
DQE defined by Persson\emph{~et~al.}\citealp{Persson2018,Persson2019,Persson2020}
is a measure of the efficiency with which an SXD converts x-ray quanta
incident upon it to a set of basis-material images. A similar approach
was used by Rajhabandary\emph{~et~al.},\citealp{Rajbhandary2017,Rajbhandary2018,Rajbhandary2020}
who defined a task-based presampling DQE in terms of the signal-to-noise
ratios (SNRs) of generalized least squares (GLS) estimates of the
amplitudes of pure sinusoidal basis materials of varying frequency.
Rajhabandary\emph{~et~al.} also evaluated the presampling DQE for
pseudo-monoenergetic imaging tasks.\citealp{Rajbhandary2017,Rajbhandary2018,Rajbhandary2020}

An important aspect of these formalisms is normalization of the frequency-dependent
SNR of an SXD by that of an ideal SXD for the same task. Persson\emph{~et~al.}\citealp{Persson2020}
defined an ideal SXD as one that does not suffer from any spectral
degradation, for example due to charge sharing caused by the finite
size charge clouds or reabsorption of characteristic photons, and
that has an impulse response function equal to a Dirac delta function,
i.e. there is no loss of spatial resolution. In addition, it was assumed
that the ideal detector has 1-keV energy bins; detectors in prototype
spectral CT systems typically have four to eight energy bins. As such,
Persson\emph{~et~al}.'s ideal detector performs a high-resolution
measurement of the x-ray spectra incident upon it, suffers no loss
of spatial information, and does not increase noise inherent in the
incident x-ray distribution. Rajhabandary\emph{~et~al.}\citealp{Rajbhandary2017,Rajbhandary2018,Rajbhandary2020}
similarly define an ideal detector but is not clear whether their
ideal detector has more energy bins than the actual detector under
study.
\begin{table*}
\caption{\label{tab:RQA}Properties of the RQA X-ray spectra.\citealp{IEC61267,IEC2003a}}

\centering{}%
\begin{tabular}{ccccc}
\hline
Spectrum & Tube Voltage {[}kV{]} & Al Filtration {[}mm{]} & Al HVL {[}mm{]} & $\mathrm{SNR_{in}}^{2}\times10^{-2}$ {[}cm$^{-2}$$\mu$Gy$^{-1}${]}\tabularnewline
\hline
RQA3 & 50 & 10.0 & 3.8 & 21759\tabularnewline
RQA4 & 60 & 16.0 & 5.4 & \tabularnewline
RQA5 & 70 & 21.0 & 6.8 & 30174\tabularnewline
RQA6 & 80 & 26.0 & 8.2 & \tabularnewline
RQA7 & 90 & 30.0 & 9.2 & 32362\tabularnewline
RQA8 & 100 & 34.0 & 10.1 & \tabularnewline
RQA9 & 120 & 40.0 & 11.6 & 31077\tabularnewline
RQA10 & 150 & 45.0 & 13.3 & \tabularnewline
\hline
\end{tabular}
\end{table*}
A limitation of these works is that the x-ray spectra that were investigated
were hardened by large quantities of water, e.g. 20\,cm. This poses
a problem for experimentation because of the large amount of scatter
that would be produced by such imaging phantoms. The presence of scatter
complicates computation of both the x-ray fluence incident on the
detector and the energy spectrum incident on the detector, both of
which are necessary to compute the performance of an ideal detector.
Another limitation is that simple closed form expressions of the task-based
SNR of ideal SXDs were not provided. This makes reproducing performance
benchmarks (i.e. the performance of the ideal detector) difficult,
frustrating a standardized approach to experimental implementation
of Persson~\emph{et~al}.'s DQE formalism.

In contrast, the International Electrotechnical Commission (IEC) has
defined a set of standardized x-ray spectra for assessing the performance
of energy-integrating detectors (EIDs).\citealp{IEC61267} These RQA
spectra (listed in Tab.~\ref{tab:RQA}) are defined in terms of nominal
tube voltages and aluminum (Al) half-value-layers (HVLs) and are produced
by hardening the spectra exiting x-ray tube windows with Al filters.
A few millimeters to a few centimeters of Al is sufficient to produce
x-ray spectra with Al HVLs similar to those encountered in a wide
range of x-ray imaging applications. These filters can be placed at
the exit window of the tube, mitigating against scatter. The resulting
spectra incident upon the detector are easily reproducible in different
laboratories.

The IEC recommends assessing detector performance for RQA spectra
in terms of the DQE:\citealp{Cunningham2000,IEC2003a}
\begin{equation}
\mathrm{DQE}\left(\uv\right)=G^{2}\mathrm{MTF}^{2}\left(\uv\right)\dfrac{W_{\mathrm{in}}\left(\uv\right)}{W_{\mathrm{out}}\left(\uv\right)}
\end{equation}
where $G$ {[}signal~cm$^{2}${]}, $\uv$ {[}cm$^{-1}${]}, $\mathrm{MTF}(\uv$),
$W_{\mathrm{in}}(\uv)$ {[}cm$^{-2}${]}, and $W_{\mathrm{out}}(\uv)$
{[}signal$^{2}$cm$^{2}${]} represent the large-area gain, two-dimensional
spatial frequency vector, modulation transfer function (MTF), noise
power spectrum (NPS) of the quanta incident upon the detector, and
measured NPS, respectively. The MTF and NPS are measured using standard
approaches,\citealp{Samei1998,Siewerdsen2002} and the incident NPS
is calculated as\citealp{IEC2003a}
\begin{equation}
W_{\mathrm{in}}\left(\uv\right)=K_{a}\mathrm{SNR}_{\mathrm{in}}^{2}=\qobar\label{eq:W_in_EID}
\end{equation}
where $K_{a}$ {[}$\mu$Gy{]} represents the air Kerma of the spectrum
incident upon the detector, $\mathrm{SNR}_{\mathrm{in}}^{2}$ {[}cm$^{-2}$$\mu$Gy$^{-1}${]}
represents the SNR of the incident quanta per unit air Kerma, and
$\qobar$ {[}cm$^{-2}${]} represents the incident x-ray fluence.
Equation~(\ref{eq:W_in_EID}) is a simple closed-form expression
for the NPS of an ideal x-ray detector, and $\mathrm{SNR}_{\mathrm{in}}^{2}$
has been tabulated for RQA x-ray spectra (as in Tab.~\ref{tab:RQA}),
providing a practical framework for standardizing the assessment of
EIDs.

In this paper, we propose using RQA x-ray spectra for assessing the
performance of SXDs, provide a precise definition of an ideal SXD,
derive mathematical expressions for the NPS of ideal SXDs for detection
and quantification tasks, including basis material decomposition and
pseudo-monoenergetic imaging, and tabulate values of the NPS of ideal
SXDs for selected tasks. The NPS values tabulated here can be used
to normalize the task-dependent DQE of SXDs in the same way that $\mathrm{SNR}_{\mathrm{in}}^{2}$
in Tab.~\ref{tab:RQA} is used to normalized the DQE of EIDs, providing
a step towards a standardized experimental framework for assessing
the performance of SXDs.

\section{Methods}

In this section, we define an ideal SXD, derive simple closed-form
expressions for the frequency-dependent SNR of ideal SXDs for detection
and quantification tasks, and describe our methodology for tabulating
the NPS for ideal SXDs for selected imaging tasks. In all cases, we
assume the properties of wide-sense stationarity are satisfied and
assume signal-known-exactly/background-known-exactly (SKE/BKE) tasks.
In what follows, a variable with a ``hat'' represents the Fourier
transform of the underlying variable, for example, $\widehat{y}$
represents the Fourier transform of $y$. A table of the most important
quantities is found in appendix \ref{sec:Table-of-Symbols} , and
to facilitate putting this work in relation to previous work in the
field this table also lists the corresponding notation in Persson\emph{~et~al.}\citealp{Persson2018}

\subsection{\label{subsec:The-Ideal-SXD}The Ideal SXD}

\begin{figure*}
\includegraphics{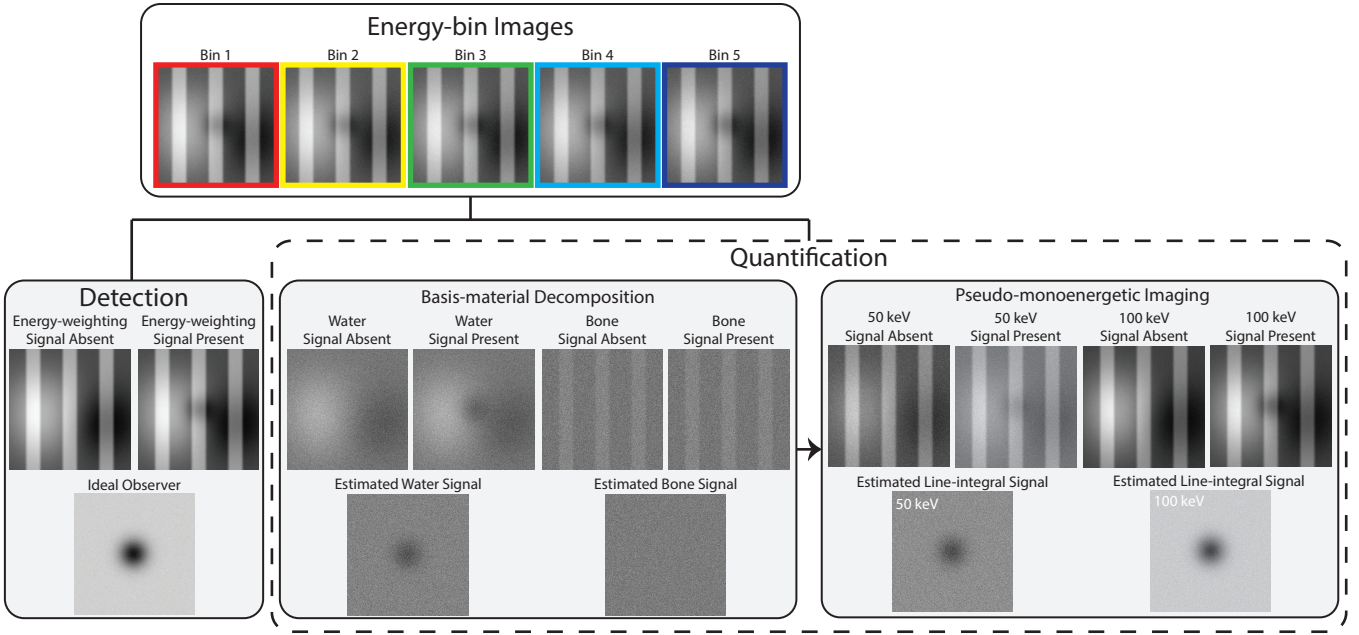}

\caption{\label{fig:Detection_vs_quantification}An illustration of detection
and quantification of a known water signal in a known water-and-bone
background. For detection, the ideal observer applies a spatio-energetic
signal template. For quantification, the image is decomposed into
water and bone basis materials, which can be used to produce pseudo-monoenergetic
images.}
\end{figure*}
Similar to Persson\emph{~et~al},\citealp{Persson2018,Persson2019,Persson2020}
we define an ideal SXD as one that does not blur the incident distribution
of x-ray quanta in the spatial domain, does not distort the spectrum
of photon energies incident upon it, does not increase the noise inherent
in the incident x-ray quanta, and has an infinite number of infinitesimal
energy bins.

With this definition, in the background region of an image, the average
number of detected photons ($\cbar$) per unit energy per unit area
is given by
\begin{equation}
\dfrac{\partial^{2}\bar{c}_{\mathrm{ideal}}}{\partial E\partial A}=\phi\left(E\right)\label{eq:Ideal_bin_counts}
\end{equation}
where $\phi\left(E\right)$ {[}keV$^{-1}$cm$^{-2}${]} represents
the fluence per unit energy incident upon the SXD. In what follows,
we assume that $\phi\left(E\right)$ does not vary with position when
the signal to be detected or quantified is absent. The corresponding
NPS per unit energy per unit area is given by
\begin{equation}
\dfrac{\partial^{2}W_{c,\mathrm{ideal}}}{\partial E\partial A}=\phi\left(E\right).\label{eq:Ideal_NPS}
\end{equation}
In what follows, we use these simple definitions to establish the
NPS of ideal SXDs for quantification and detection tasks.

\subsection{Basis-material Decomposition and Quantification Tasks}

The goal of basis material decomposition is to use the spectral information
provided by an SXD to decompose the line-integral of x-ray attenuation
into two or more basis-material images, as illustrated in Fig.~\ref{fig:Detection_vs_quantification}.
We consider the task of estimating the signal difference due to a
perturbation ($\Delta a$) in the area density ($a$) {[}g/cm$^{2}${]}
of one or more basis materials. This task is illustrated in Fig.~\ref{fig:Detection_vs_quantification}
for a non-uniform (but known) background. Unlike the task in Fig.~\ref{fig:Detection_vs_quantification},
which has a non-stationary background, we assume the background is
uniform and stationary. In this case, the average x-ray fluence per
unit energy is uniform over the detector when the signal is absent,
but there will be stochastic fluctuations from one region to another.

In the Fourier domain, the average difference in the line integral
of x-ray attenuation between signal-present and signal-absent conditions
is given by
\begin{equation}
\widehat{\Delta l}\left(\uv\right)=\sum_{b=1}^{B}\tfrac{\mu}{\rho}_{b}\left(E\right)\widehat{\Delta a}_{b}\left(\uv\right)\label{eq:Line_integral}
\end{equation}
where $\widehat{\Delta a}_{b}(\uv)$ {[}g{]} represents the Fourier
transform of the signal difference of basis material $b$, $\mu/\rho$
{[}cm$^{2}$/g{]} represents the mass attenuation coefficient, and
$B$ represents the number of basis materials. At this point we keep
our derivation generic to any number of basis materials. We focus
on particular sets of basis materials in Sec.~\ref{subsec:Calculations}.

The average signal differences ($\widehat{\Delta a}_{b}(\uv)$) are
estimated from noisy SXD data. We can therefore define an NPS for
each basis material image in addition to a cross NPS that quantifies
noise correlations between basis materials, as described below.

\subsubsection{Basis-material NPS for the Ideal SXD}

To derive a closed-form expression for the NPS of an ideal SXD, we
make the following assumptions about the elements of the vector $\widehat{\Delta\mathbf{a}}$:
(1) they are real and (2) their power is concentrated within the Nyquist
region defined by the pixel pitch of the detector. The first assumption
simplifies our mathematical analysis and the corresponding notation
without affecting the mathematical results, which are ultimately independent
of $\widehat{\Delta\mathbf{a}}$. The second assumption ensures that
aliasing is negligible, which is satisfied when the signal to be detected
is larger than a few detector elements in both directions, is common
in the field and enables utilizing linear systems theory.

Similar to Rajhabandary \emph{et~al.}\citep{Rajbhandary2020} and
Persson~\emph{et~al.},\citealp{Persson2018,Persson2020} we are
interested in establishing an upper limit of image quality. To this
end, we assume that $\widehat{\Delta\mathbf{a}}$ is estimated using
an unbiased estimator that achieves the Cramer-Rao lower bound. We
therefore assume the generalized least squares (GLS) estimator to
the linearized basis material problem, which, with the assumptions
listed above, can be posed as
\begin{equation}
\widehat{\Delta\mathbf{L}}=\widehat{\mathbf{M}}\widehat{\Delta\mathbf{a}}+\widehat{\boldsymbol{\varepsilon}}\label{eq:GLS_problem}
\end{equation}
where $\widehat{\boldsymbol{\varepsilon}}$ represents measurement
noise which is assumed to be additive, and
\begin{equation}
\widehat{\Delta L}_{i}=-FT\left[\log\dfrac{c_{i,0}+\Delta c_{i}}{\bar{c}_{i,0}}\right]\approx-\dfrac{\widehat{\Delta c}_{i}}{\bar{c}_{i,0}}\label{eq:log_transform}
\end{equation}
where $\widehat{\Delta c}_{i}$ represents the signal difference in
energy bin $i$ due to the deviation $\widehat{\Delta\mathbf{a}}$
and $\bar{c}_{i,0}$ is the average number of photons per element
in the background region of the image formed from energy bin $i$.
The matrix $\widehat{\mathbf{M}}$ is a transformation matrix that
maps from the space of log-normalised energy bins to the space of
basis materials. In general, $\widehat{\mathbf{M}}$ is frequency
dependent, accounting for the spatial resolution of each energy bin
in addition to spectral distortions that lead to photons counted in
the wrong energy bin. For an ideal detector, the MTF of each energy
bin is unity for all spatial frequencies and every incident photon
is counted in the correct energy bin. When the linear approximation
in Eq.~(\ref{eq:GLS_problem}) is satisfied, and for a detector with
finite energy bins but otherwise ideal, an unbiased estimate is obtained
when the matrix $\widehat{\mathbf{M}}$ has elements given by
\begin{equation}
\widehat{M}_{i,b}=\dfrac{{\displaystyle \int_{E_{i}}^{E_{i}+\Delta E_{i}}\tfrac{\mu}{\rho}_{b}\left(E\right)\phi\left(E\right)\mathrm{d}E}}{{\displaystyle \int_{E_{i}}^{E_{i}+\Delta E_{i}}\phi\left(E\right)\mathrm{d}E}}\label{eq:Transformation_Matrix}
\end{equation}
where $E_{i}$ and $E_{i}+\Delta E_{i}$ {[}keV{]} represent the lower
and upper limits of energy bin $i$, and (as described above) $\phi\left(E\right)$
{[}cm$^{-2}$keV$^{-1}${]} represents the fluence per unit energy.

For the GLS estimator, the frequency-dependent covariance matrix of
the estimate of $\widehat{\Delta\mathbf{a}}$ is given by
\begin{equation}
\mathbf{W}_{\mathbf{a}}=\left(\widehat{\mathbf{M}}^{T}\mathbf{W}_{L}^{-1}\widehat{\mathbf{M}}\right)^{-1}\label{eq:GLS_Covariance}
\end{equation}
where $\mathbf{W}_{L}$ represents the covariance matrix of the measurements
$\widehat{\Delta\mathbf{L}}$ when the signal is absent. The $i^{\mathrm{th}}$
diagonal element of $\mathbf{W}_{L}$ represents the NPS of $\Delta L_{i}$;
element $i,j$ of $\mathbf{W}_{L}$ represents the cross NPS between
$\Delta L_{i}$ and $\Delta L_{j}$ under signal-absent conditions.
In all cases, for both ideal and non-ideal detectors, we assume that
$\mathbf{W}_{L}$ is real-valued and symmetric, which makes it Hermitian.
These assumptions are likely satisfied in real SXDs in which physical
processes that result in correlated noise are isotropic in the detector
plane, resulting in spatial correlations that are symmetric to reflections
about the two axes that define the detector plane.

Combining Eqs.~(\ref{eq:Ideal_bin_counts}), (\ref{eq:Ideal_NPS})
and (\ref{eq:log_transform}) for a detector with energy bins of finite
width but otherwise ideal, the elements of $\mathbf{W}_{L}$ are given
by
\begin{equation}
\left[\mathbf{W}_{L}\right]_{i,j}=\delta_{ij}\left(\int_{E_{i}}^{E_{i}+\Delta E_{i}}\phi\left(E\right)\mathrm{d}E\right)^{-1}\,\,\,\text{finite bin widths}\label{eq:W_L_finite}
\end{equation}
where $\delta_{ij}$ represents the Kronecker delta function.

\begin{figure}
\begin{centering}
\includegraphics{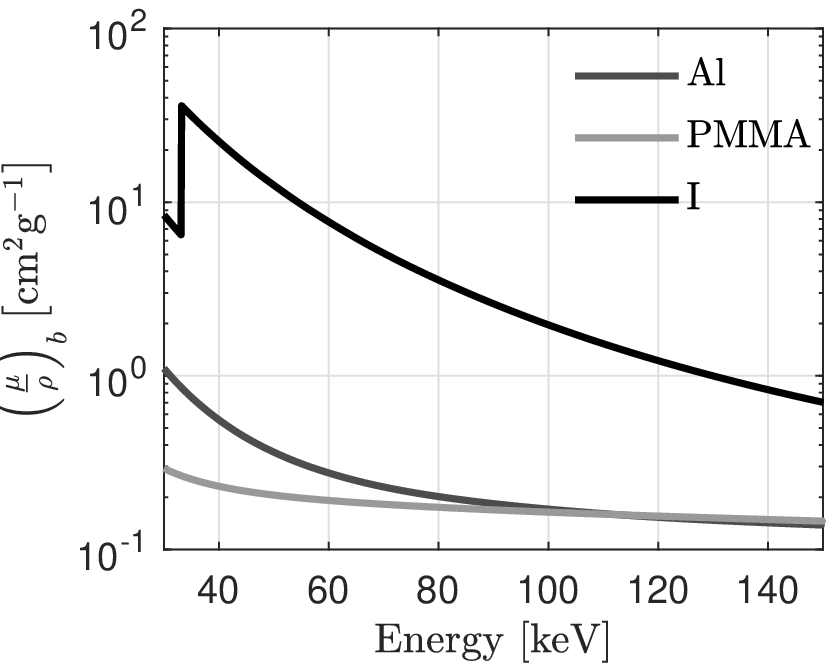}
\par\end{centering}
\begin{centering}
\includegraphics{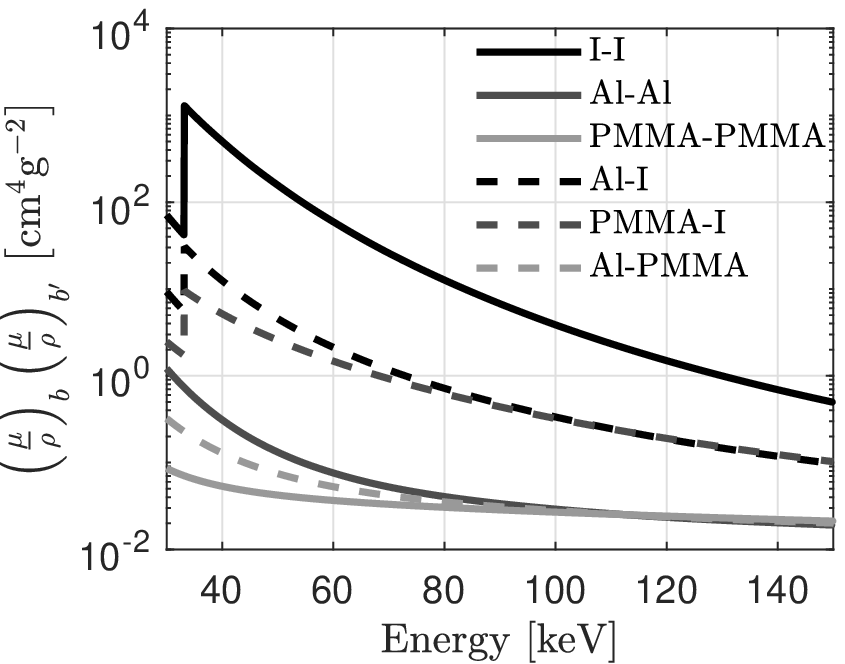}
\par\end{centering}
\caption{\label{fig:Cross_terms}The top plot shows the mass-attenuation coefficients
of Al, PMMA and Iodine. The bottom plot shows the products of mass-attenuation
coefficients, which are relevant for the inner product in Eq.~(\ref{eq:correlation_matrix}).}
\end{figure}

In the limit of an infinite number of energy bins of infinitesimal
width, Eqs.~(\ref{eq:Transformation_Matrix}) and (\ref{eq:W_L_finite})
become
\begin{equation}
M_{i,b}\approx\dfrac{\mu}{\rho}_{b}\left(E_{i}\right)
\end{equation}
and
\begin{equation}
\left[\mathbf{W}_{L}\right]_{i,j}\approx\dfrac{\delta_{ij}}{\phi\left(E_{i}\right)\Delta E_{i}}.
\end{equation}
Combining these results with Eq.~(\ref{eq:GLS_Covariance}) yields
\begin{align}
\lim_{\Delta E\rightarrow0}\left[\mathbf{W}_{\mathbf{a}}^{-1}\right]_{b,b^{\prime}} & =\qobar\left\langle \tfrac{\mu}{\rho}_{b},\tfrac{\mu}{\rho}_{b^{\prime}}\right\rangle _{\phi}\label{eq:Inverse_Covariance_ideal}
\end{align}
where $\qobar=\int_{0}^{\infty}\phi(E)\mathrm{d}E$ and $\left\langle (\mu/\rho)_{b},\right.\left.(\mu/\rho)_{b^{\prime}}\right\rangle _{\phi}$
represents the inner product of $(\mu/\rho)_{b}$ and $(\mu/\rho)_{b^{\prime}}$
with respect to $\phi(E)/\qobar$:
\begin{equation}
\left\langle \tfrac{\mu}{\rho}_{b},\tfrac{\mu}{\rho}_{b^{\prime}}\right\rangle _{\phi}=\dfrac{1}{\qobar}{\displaystyle \int_{0}^{\infty}\tfrac{\mu}{\rho}_{b}\left(E\right)\tfrac{\mu}{\rho}_{b^{\prime}}\left(E\right)\phi\left(E\right)\mathrm{d}E}.\label{eq:correlation_matrix}
\end{equation}
We therefore have
\begin{equation}
\mathbf{W}_{\mathbf{a},\mathrm{ideal}}=\dfrac{1}{\qobar}\mathbf{M}_{2}^{-1}\label{eq:ideal_basis_noise_power}
\end{equation}
where element $b,b^{\prime}$ of the matrix $\mathbf{M}_{2}$ is given
by Eq.~(\ref{eq:correlation_matrix}) and $\mathbf{W}_{\mathbf{a},\mathrm{ideal}}$
represents the frequency-dependent covariance matrix of basis materials
for the ideal detector defined in Sec.~\ref{subsec:The-Ideal-SXD}.
The matrix $\mathbf{M}_{2}$ has units of cm$^{4}$/g$^{2}$ and the
fluence has units of cm$^{-2}$ which yields units of g$^{2}$/cm$^{2}$
for the elements of $\mathbf{W}_{\mathbf{a},\mathrm{ideal}}$, as
necessary. The $b^{\mathrm{th}}$ diagonal element of $\mathbf{W}_{\mathbf{a},\mathrm{ideal}}$
represents the NPS of basis material $b$; element $b,b^{\prime}$
the cross NPS between basis materials $b$ and $b^{\prime}$. The
cross NPS between materials $b$ and $b^{\prime}$ is in general non-zero
for an ideal SXD because the mass attenuation coefficients of any
two basis materials do not form an orthogonal basis set with respect
to the x-ray spectrum $\phi_{0}(E)$, i.e. the right side of Eq.~(\ref{eq:correlation_matrix})
is non-zero for $b\ne b^{\prime}$. For example, the product of the
mass attenuation coefficients of Al, poly-methyl-methacrylate (PMMA)
and iodine, which are common basis materials, are shown in Fig.~\ref{fig:Cross_terms}.

Equation~(\ref{eq:ideal_basis_noise_power}) is a novel theoretical
contribution of this work. It is useful because the matrix $\mathbf{M}_{2}$
(or its inverse) can be tabulated for standardized x-ray spectra for
basis sets of relevance, as described in Sec.~\ref{subsec:Calculations}.

\subsubsection{DQE for Quantification Tasks}

For an ideal detector, the frequency-dependent squared SNR of the
unbiased GLS estimate of the area density of basis material $b$ is
given by
\begin{equation}
\mathrm{SNR}_{b,\mathrm{ideal}}^{2}=\dfrac{\widehat{\Delta a}_{b}^{2}}{\left[\mathbf{W}_{\mathbf{a},\mathrm{ideal}}\right]_{b,b}}.\label{eq:SNR_ideal}
\end{equation}
If we similarily assume an unbiased GLS estimate for a non-ideal detector,
the resulting SNR given is by
\begin{equation}
\mathrm{SNR}_{b}^{2}=\dfrac{\widehat{\Delta a}_{b}^{2}}{\left[\mathbf{W}_{\mathbf{a}}\right]_{b,b}}\label{eq:SNR_nonideal}
\end{equation}
where $\mathbf{W}_{\mathbf{a}}$ is the non-ideal counterpart to $\mathbf{W}_{\mathbf{a},\mathrm{ideal}}$,
and is given by Eq.~(\ref{eq:GLS_Covariance}) with $\widehat{\mathbf{M}}$
given by
\begin{equation}
\widehat{M}_{i,b}=\dfrac{{\displaystyle \int_{0}^{\infty}\tfrac{\mu}{\rho}_{b}\left(E\right)\phi\left(E\right)T_{i}\left(\uv,E\right)\mathrm{d}E}}{{\displaystyle \int_{0}^{\infty}\phi\left(E\right)T_{i}\left(0,E\right)\mathrm{d}E}}\,\,\,\mathrm{non\,ideal}\label{eq:Realistic_M}
\end{equation}
where $T_{i}(\uv,E)$ {[}cm$^{2}${]} represents the energy-dependent
characteristic transfer function of energy bin $i$. The characteristic
transfer function is the Fourier transform of the impulse response
function.\citealp{Cunningham2000} The impulse response function of
energy bin $i$ of an SXD is equal to the number of photons detected
in energy bin $i$ of a detector element centered at position $\mathbf{r}$
given a photon incident at the origin.\citealp{Tanguay2018} In writing
Eq.~(\ref{eq:Realistic_M}) we have assumed that $T_{i}(\uv,E)$
is real, which is satisfied if the impulse response function of each
energy bin is real and symmetric. Element $i,b$ of $\widehat{\mathbf{M}}$
represents the frequency-dependent response of energy $i$ to the
spectrum $\phi$ weighted by the mass-attenuation coefficient of basis
material $b$. It could be measured using thin filters comprised of
the individual basis materials and a slanted edge.

Taking the ratio of Eq.~(\ref{eq:SNR_nonideal}) to Eq.~(\ref{eq:SNR_ideal})
yields the DQE for quantification tasks:
\begin{equation}
\mathrm{DQE}_{b}^{Q}\left(\uv\right)=\dfrac{\left[\mathbf{M}_{2}^{-1}\right]_{b,b}}{\qobar\left[\left(\widehat{\mathbf{M}}^{T}\mathbf{W}_{L}^{-1}\widehat{\mathbf{M}}\right)^{-1}\right]_{b,b}}\label{eq:DQE}
\end{equation}
where the superscript $Q$ indicates that the task is quantification.
This result shows that knowledge of $\mathbf{M}_{2}^{-1}$ is required
to compute the DQE for quantification tasks. We provide tables of
the elements of $\mathbf{M}_{2}^{-1}$ for selected tasks and standardized
spectra, as described in Sec.~\ref{subsec:Calculations}.

\subsubsection{Pseudo-monoenergetic Imaging}

One purpose of basis material decomposition is to produce pseudo monoenergetic
images by estimating the energy-dependent line-integral of x-ray attenuation
in Eq.~(\ref{eq:Line_integral}). Assuming an unbiased GLS estimate
of $\widehat{\Delta\mathbf{a}}$ and ideal detectors with an infinite
number of infinitesimal energy bins, the noise power of a pseudo monoenergetic
image is given by
\begin{equation}
W_{l,\mathrm{ideal}}=\dfrac{1}{\qobar}\bm{\mu}^{T}\mathbf{M}_{2}^{-1}\bm{\mu}\label{eq:Line_integral_NPS}
\end{equation}
 where $\bm{\mu}$ represents a vector of basis material mass attenuation
coefficients. Note that although not explicit, $W_{l,\mathrm{ideal}}$
is a function of energy through $\boldsymbol{\mu}$, the elements
of which represent the energy-dependent mass-attenuation coefficients
of basis materials.

Similar to the basis material SNR, we can normalize the SNR of the
unbiased GLS estimate of $\widehat{\Delta l}$ for non-ideal detectors
by that of ideal detectors to provide a DQE:
\begin{equation}
\mathrm{DQE}^{E}\left(\fv\right)=\dfrac{1}{\qobar}\dfrac{\bm{\mu}^{T}\mathbf{M}_{2}^{-1}\bm{\mu}}{\bm{\mu}^{T}\left(\widehat{\mathbf{M}}^{T}\mathbf{W}_{L}^{-1}\widehat{\mathbf{M}}\right)^{-1}\bm{\mu}}\label{eq:DQE_E}
\end{equation}
where $(\widehat{\mathbf{M}}^{T}\mathbf{W}_{L}^{-1}\widehat{\mathbf{M}})^{-1}=\mathbf{W}_{a}$
represents the basis material noise power matrix for non-ideal detectors
and the superscript $E$ indicates mono-energetic imaging. This ratio
is equivalent to the definition of the DQE for pseudo-monoenergetic
tasks described by Rahjabandary~et~al.\citealp{Rajbhandary2020}
Again, providing a normalized metric of image quality requires knowledge
of $\mathbf{M}_{2}^{-1}$.

\subsection{Detection Tasks}

In a detection task, the ideal observer uses their knowledge of the
signal and noise properties of the energy-bin images to whiten the
image data by decorrelating it in the spatial and energy domains and
then applying the signal template to the data.\citealp{ImageScience,Persson2018}
In the energy domain, applying the signal template takes the form
of energy weighting;\citealp{Tapiovaara1985,Persson2018} in the spatial
domain, the known shape of the signal is used as a window through
which the image is viewed, as illustrated in Fig.~\ref{fig:Detection_vs_quantification}.
For an SXD, the squared SNR of the ideal-observer is given by\citealp{Persson2018,Persson2019}
\begin{equation}
\mathrm{SNR}^{2}=\int_{\mathrm{Nyq}}\widehat{\Delta\boldsymbol{\mathrm{c}}}^{T}\mathbf{W}^{-1}\widehat{\Delta\boldsymbol{\mathrm{c}}}\mathrm{d}\fv\label{eq:Detectability}
\end{equation}
where $\widehat{\Delta\boldsymbol{\mathrm{c}}}$ {[}cm$^{2}${]} represents
a vector with elements equal to the signal difference of each energy
bin:
\begin{equation}
\widehat{\Delta c}_{i}\left(\fv\right)=\int_{0}^{\infty}T_{i}\left(\fv,E\right)\widehat{\Delta\Phi}\left(\fv,E\right)\mathrm{d}E
\end{equation}
where, for small perturbations, $\Delta\Phi\left(\fv,E\right)$ {[}keV$^{-1}${]}
is given by
\begin{equation}
\widehat{\Delta\Phi}\left(\fv,E\right)=\phi\left(E\right)\sum_{b=1}^{B}\tfrac{\mu}{\rho}_{b}\left(E\right)\widehat{\Delta a}_{b}.
\end{equation}
For a detector with energy bins of finite width but otherwise ideal,
$\Delta S_{i}\left(\fv\right)$ becomes
\[
\widehat{\Delta c}_{i}\left(\fv\right)=\bar{c}_{i,0}\sum_{b=1}^{B}\widehat{M}_{i,b}\widehat{\Delta a}_{b}
\]
where $\widehat{M}_{i,b}$ is given by Eq.~(\ref{eq:Transformation_Matrix}).
We also note that $W_{ij}=\delta_{ij}\bar{c}_{i}A$ for a detector
with energy bins of finite width but otherwise ideal. The ideal-observer
SNR then becomes
\begin{equation}
\mathrm{SNR}^{2}=\int\sum_{b=1}^{B}\sum_{b^{\prime}=1}^{B}\sum_{i=1}^{N}\bar{c}_{i,0}\tfrac{\mu}{\rho}_{i,b}\tfrac{\mu}{\rho}_{i,b^{\prime}}\widehat{\Delta a}_{b}\widehat{\Delta a}_{b^{\prime}}\mathrm{d}\fv.
\end{equation}
In the limit of an infinite number of infinitesimal energy bins
\begin{equation}
\mathrm{SNR}_{\mathrm{ideal}}^{2}=\qobar\int\widehat{\Delta\mathbf{a}}^{T}\mathbf{M}_{2}\widehat{\Delta\mathbf{a}}\,\mathrm{d}\fv\label{eq:Ideal_SNR_Detection_M2}
\end{equation}
where the elements of $\mathbf{M}_{2}$ are given by Eq.~(\ref{eq:correlation_matrix}).
We can alternatively express the SNR in terms of $\mathbf{W}_{\mathbf{a},\mathrm{ideal}}$
by making use of Eq.~(\ref{eq:ideal_basis_noise_power}):
\begin{equation}
\mathrm{SNR}_{\mathrm{ideal}}^{2}=\int\widehat{\Delta\mathbf{a}}^{T}\mathbf{W}_{\mathbf{a},\mathrm{ideal}}^{-1}\widehat{\Delta\mathbf{a}}\,\mathrm{d}\fv.\label{eq:Ideal_SNR_Detection}
\end{equation}
Equations~(\ref{eq:Ideal_SNR_Detection_M2}) and (\ref{eq:Ideal_SNR_Detection})
show that the ideal observer SNR for ideal detectors is determined
by the matrix of basis material noise power for the ideal detector,
which is in turn determined by the matrix of inner products $\mathbf{M}_{2}$.

\subsubsection{DQE for Single-material Detection Tasks}

We consider the case when the signal difference is due to the perturbation
of a single basis material. In this case, the ideal ideal-observer
SNR is given by
\begin{equation}
\mathrm{SNR}_{\mathrm{ideal},b}^{2}=\int\widehat{\Delta a_{b}}^{2}\left[\mathbf{W}_{\mathbf{a},\mathrm{ideal}}^{-1}\right]_{b,b}\,\mathrm{d}\fv.\label{eq:Ideal_Detection_basis}
\end{equation}
The non-ideal ideal-observer SNR can be shown to be given by
\begin{align}
\mathrm{SNR}_{b}^{2} & =\int_{\mathrm{Nyq}}\widehat{\Delta a}_{b}^{2}\,\widehat{\boldsymbol{\mathrm{M}}}_{b}^{T}\mathbf{W}_{L}^{-1}\widehat{\boldsymbol{\mathrm{M}}}_{b}\mathrm{d}\fv\label{eq:Actual_detection_basis}
\end{align}
where $\widehat{\mathbf{M}}_{b}$ represents column $b$ of the matrix
with elements given by Eq.~(\ref{eq:Realistic_M}). In writing Eq.~(\ref{eq:Actual_detection_basis})
we have changed the basis set from that of energy bin counts to that
of the basis materials. The integrands of Eqs.~(\ref{eq:Ideal_Detection_basis})
and (\ref{eq:Actual_detection_basis}) represent frequency-dependent
SNRs; dividing the latter by the former yields the DQE of basis material
$b$ for detection tasks:
\begin{equation}
\mathrm{DQE}_{b}^{D}\left(\fv\right)=\dfrac{\widehat{\boldsymbol{\mathrm{M}}}_{b}^{T}\mathbf{W}_{L}^{-1}\widehat{\boldsymbol{\mathrm{M}}}_{b}}{\qobar\left[\mathbf{M_{2}}\right]_{b,b}}\label{eq:DQE_D}
\end{equation}
where the superscript $D$ indicates ``detection.'' Again, we see
that computing the frequency-dependent DQE requires knowledge of the
matrix $\mathbf{M}_{2}$.

\subsection{\label{subsec:Calculations}Calculation of $\mathbf{W}_{\mathbf{a},\mathrm{ideal}}$}

We calculated the matrix of inner products ($\mathbf{M}_{2}$) and
its inverse for RQA x-ray spectra (listed in Tab.~\ref{tab:RQA})
and selected sets of basis materials, as described below.

\subsubsection{X-ray Spectrum}

We simulated poly-energetic x-ray spectra using the Tucker and Barnes
algorithm\citealp{Tucker1991} implemented using an in-house MATLAB
script. We simulated a tungsten (W) anode, a target angle of 10\,degrees
and inherent filtration by 2.38\,mm of pyrex, 2.66\,mm of Lexan,
3.06\,mm of oil and 1.5\,mm of Al. We simulated the number of photons
in 0.1-keV energy bins for energies ranging from 20\,keV to the maximum
energy of the x-ray spectrum. We then used the Lambert-Beer Law to
simulate hardening of x-ray spectra by the additional Al filtration
listed in Tab.~\ref{tab:RQA}. The HVL was then verified by comparing
the HVLs of the modeled spectra with the desired HVLs in Tab.~\ref{tab:RQA}.
The thickness of the added Al filtration was increased or decreased
until the modeled HVL was within 0.25\% of the desired HVL. For the
HVL verification, we calculated the air Kerma using the mass energy
transfer coefficient of air in the NIST database.

\subsubsection{Basis Materials}

The elements of the matrix $\mathbf{M}_{2}$ are functions of the
mass attenuation coefficients of the set of basis materials. Clinically
relevant basis materials include soft-tissue and bone, in addition
to contrast agents such as iodine and gadolinium. Because of their
similar mass attenuation coefficients, water or PMMA, i.e. acrylic,
are often used in place of soft-tissue in x-ray imaging experiments;
Al is often used in place of bone. We therefore computed $\mathbf{M}_{2}$
for a basis set consisting of PMMA and Al, which are most amenable
to experimentation. To accommodate contrast-enhanced imaging, we also
computed $\mathbf{M}_{2}$ for basis sets consisting of PMMA, Al and
either iodine (I) or gadolinium (Gd).

The mass attenuation coefficients of basis materials were combined
with the x-ray spectra described in the preceding section to compute
the inner product in Eq.~(\ref{eq:Inverse_Covariance_ideal}). A
Riemann sum with 0.1-keV energy bins was used to numerically compute
the integral over the energy domain.

\subsubsection{Normalization of $\mathbf{W}_{\mathbf{a},\text{ideal}}$}

The spectra and basis materials from the preceding section were used
to compute $\mathbf{M}_{2}^{-1}$. To be consistent with the IEC,
we normalized $\mathbf{M}_{2}^{-1}$ such that it can be converted
to the ideal NPS for a given fluence using a measurement of the air
Kerma which, unlike the fluence, is directly accessible experimentally.
To this end, $\mathbf{W}_{\mathbf{a},\text{ideal}}$ can be expressed
as
\begin{equation}
\mathbf{W}_{\mathbf{a},\text{ideal}}=\dfrac{1}{K_{a}}\left(\dfrac{\partial K_{a}}{\partial q}\mathbf{M}_{2}^{-1}\right)\label{eq:W_ideal_practical}
\end{equation}
where $K_{a}$ {[}$\mu$Gy{]} and $\partial K_{a}/\partial q$ {[}$\mu$Gy~cm$^{2}${]}
represent the air Kerma and the air Kerma per unit fluence, respectively.
We tabulated the term in parenthesis in Eq.~(\ref{eq:W_ideal_practical})
in units of g$^{2}$cm$^{-2}$$\mu$Gy. In practice, calculating $\mathbf{W}_{\mathbf{a},\text{ideal}}$
would be achieved by dividing the tabulated values of $(\partial K_{a}/\partial q)\mathbf{M}_{2}^{-1}$
by the air Kerma used to measure the actual NPS of the SXD under study.
We present some example calculations of $(\partial K_{a}/\partial q)\mathbf{M}_{2}^{-1}$
in the results section; tables of $(\partial K_{a}/\partial q)\mathbf{M}_{2}^{-1}$
values for the RQA spectra and basis sets described above are given
in the Appendix.

\begin{figure}
\begin{centering}
\includegraphics{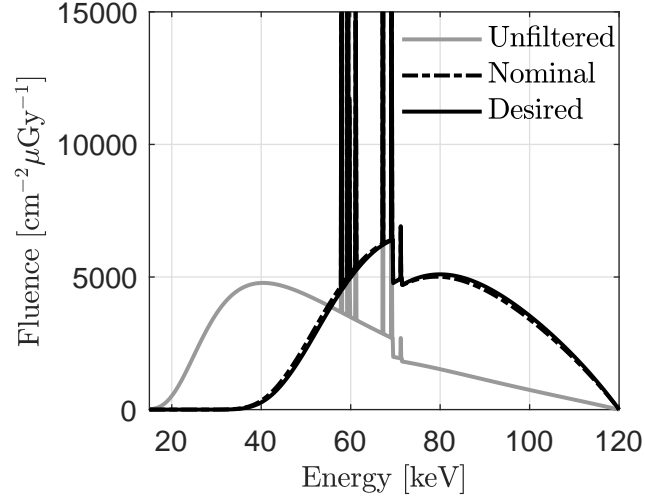}
\par\end{centering}
\caption{\label{fig:Spectrum_methods}Models of the x-ray spectra used in this
work. The nominal and desired spectra are those hardened using the
nominal added Al thickness and that which yields a HVL within 0.25\%
of the HVL prescribed by the IEC.}
\end{figure}
Equation~(\ref{eq:W_ideal_practical}) is useful for normalizing
$\mathbf{M}_{2}^{-1}$ for calculation of the DQE, for which the fluence
incident on the detector is relevant and obtained from a measurement
of air Kerma and tabulated values of $\partial K_{a}/\partial q$.
As will be shown in the Results, the resulting noise power for a fixed
air Kerma incident on the detector increases with tube voltage, apparently
suggesting that lower tube voltages are advantageous. This is somewhat
non-intuitive because the fraction of photons transmitted through
a patient increases with increasing tube voltage, resulting in more
photons reaching the detector for a given patient dose, potentially
leading to lower noise per unit patient dose for higher-energy spectra.
As such, for the purpose of illustration, we also normalized $\mathbf{M}_{2}^{-1}$
by the air Kerma incident on the patient, which we approximated as
a 20-cm-thick water slab. To this end, $\mathbf{W}_{\mathbf{a},\text{ideal}}$
was expressed as
\begin{equation}
\mathbf{W}_{\mathbf{a},\text{ideal}}=\dfrac{1}{K_{a,\mathrm{entrance}}}\left(\dfrac{\partial K_{a,\mathrm{entrance}}}{\partial q}\mathbf{M}_{2}^{-1}\right)\label{eq:W_ideal_illustration}
\end{equation}
where $K_{a,\mathrm{entrance}}$ {[}$\mu$Gy{]} and $\partial K_{a,\mathrm{entrance}}/\partial q$
{[}$\mu$Gy~cm$^{2}${]} represent the air Kerma incident on the
patient and the entrance air Kerma per unit fluence incident on the
detector, respectively.
\begin{figure}
\begin{centering}
\includegraphics{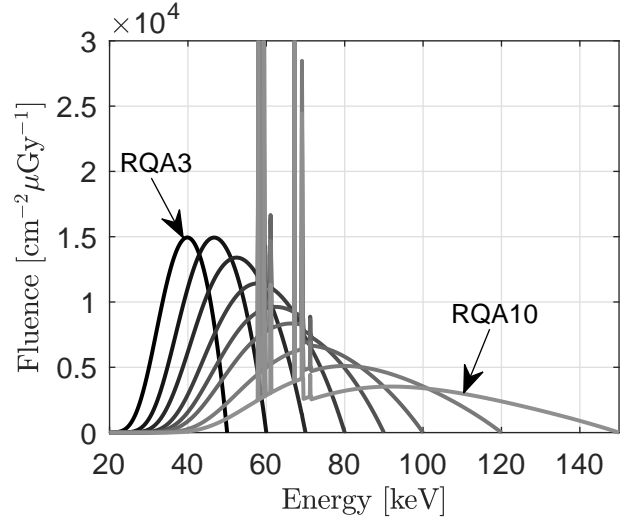}
\par\end{centering}
\caption{\label{fig:RQA-series-x-ray-spectra.}X-ray spectra used to calculate
the NPS of ideal SXDs.}
\end{figure}
\begin{figure*}
\includegraphics{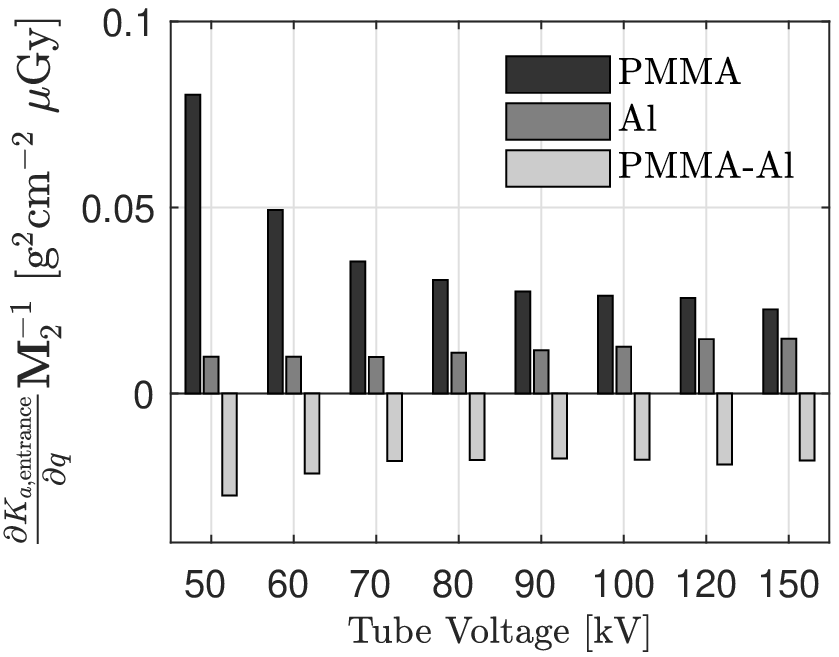}\includegraphics{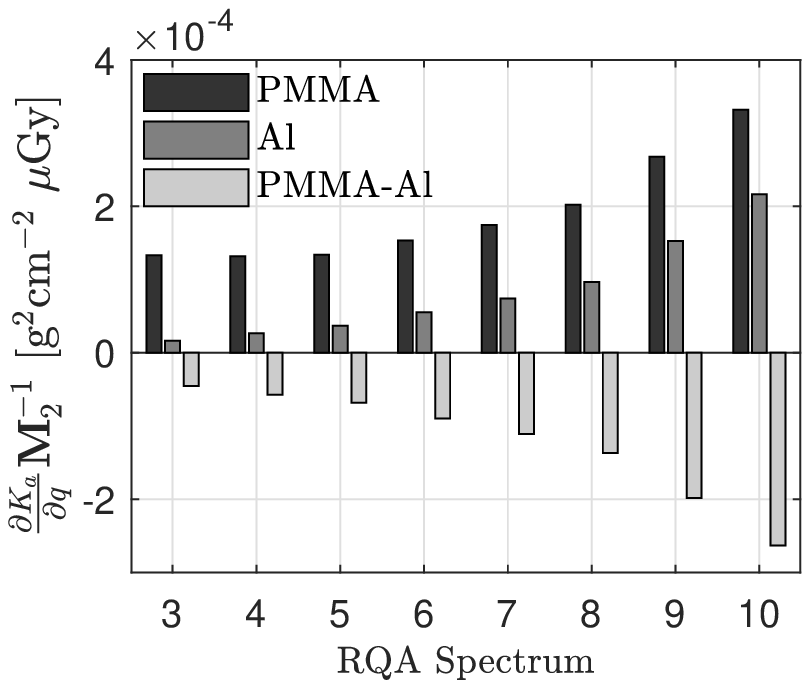}

\caption{\label{fig:Bar-plots-PMMA-Al}Bar plots of the noise power of basis-material
images of ideal SXDs for a two-material decomposition with PMMA and
Al basis materials. The variables $K_{a,\mathrm{entrance}}$, $K_{a}$
and $q$ represent the entrance air Kerma (i.e. that incident on a
patient), the air Kerma incident on the detector and the x-ray fluence
incident on the detector respectively.}
\end{figure*}

\section{Results}

\subsection{X-ray Spectra}

\begin{table*}
\caption{\label{tab:Actual_RQA}Properties of RQA x-ray spectra. The nominal
tube voltages, nominal Al filtration, and desired HVLs are those prescribed
by the IEC.\citealp{IEC61267} The nominal HVL and fluence per unit
air Kerma are those calculated using the nominal tube voltages and
nominal Al filtrations for x-ray spectra modeled using the Tucker~and~Barnes
algorithm.\citealp{Tucker1991} The actual Al filtrations, HVLs and
fluences per unit air Kerma are those used to calculate $(\partial K_{a}/\partial q)\mathbf{M}_{2}^{-1}$.}

\centering{}%
\begin{tabular}{cc|cc|ccc|cc}
\hline
 &  & \multicolumn{2}{c|}{Al Filtration {[}mm{]}} & \multicolumn{3}{c|}{ Al HVL {[}mm{]}} & \multicolumn{2}{c}{ $\partial q/\partial K_{a}\times10^{-2}$ {[}cm$^{-2}$$\mu$Gy$^{-1}${]}}\tabularnewline
Spectrum & Tube Voltage {[}kV{]} & Nominal  & Actual & Desired & Nominal & Actual & Nominal & Actual\tabularnewline
\hline
RQA3 & 50 & 10.0 & 10.3 & 3.80 & 3.77 & 3.80 & 22312 & 22408\tabularnewline
RQA4 & 60 & 16.0 & 16.9 & 5.40 & 5.33 & 5.41 & 27887 & 28127\tabularnewline
RQA5 & 70 & 21.0 & 21.8 & 6.80 & 6.74 & 6.80 & 31368 & 31515\tabularnewline
RQA6 & 80 & 26.0 & 27.5 & 8.20 & 8.09 & 8.20 & 33324 & 33484\tabularnewline
RQA7 & 90 & 30.0 & 31.0 & 9.20 & 9.15 & 9.21 & 33894 & 33950\tabularnewline
RQA8 & 100 & 34.0 & 34.6 & 10.1 & 10.0 & 10.1 & 33685 & 33696\tabularnewline
RQA9 & 120 & 40.0 & 42.5 & 11.6 & 11.5 & 11.6 & 32024 & 31948\tabularnewline
RQA10 & 150 & 45.0 & 48.0 & 13.3 & 13.2 & 13.3 & 28630 & 28427\tabularnewline
\hline
\end{tabular}
\end{table*}
Table~\ref{tab:Actual_RQA} summarizes the x-ray spectra used to
calculate $\mathbf{M}_{2}$. The nominal tube voltages, nominal Al
filtration and desired HVLs are those prescribed by the IEC.\citealp{IEC61267}
The nominal HVLs and fluences per unit air Kerma are those calculated
using the nominal tube voltages and nominal Al filtration for x-ray
spectra modeled using the Tucker~and~Barnes algorithm.\citealp{Tucker1991}
Figure~\ref{fig:Spectrum_methods} shows an example of an unfiltered
120-kV spectrum, a 120-kV spectrum hardened by the nominal Al thickness
for an RQA9 spectrum, and the desired RQA9 spectrum.

The actual Al filtration, HVLs and fluences per unit air Kerma are
those used to calculate $(\partial K_{a}/\partial q)\mathbf{M}_{2}^{-1}$.
In all cases, the Al thickness required to achieve the desired Al
HVL was greater than the nominal thickness prescribed by the IEC,
indicating that our implementation of the Tucker and Barnes algorithm\citealp{Tucker1991}
produces spectra slightly softer than those that the IEC used to determine
the nominal Al thicknesses in Tab.~\ref{tab:RQA}. Also, the fluences
per unit air Kerma calculated using the nominal Al thickness and the
actual Al thickness exceed those reported by the IEC\citealp{IEC2003a}
by 3\% to 5\%. The spectra calculated using the parameters in Tab.~\ref{tab:Actual_RQA}
are shown in Fig.~\ref{fig:RQA-series-x-ray-spectra.}.

\subsection{Basis-material NPS ($\mathbf{W}_{\mathbf{a},\text{ideal}}$)}

\begin{figure*}
\includegraphics{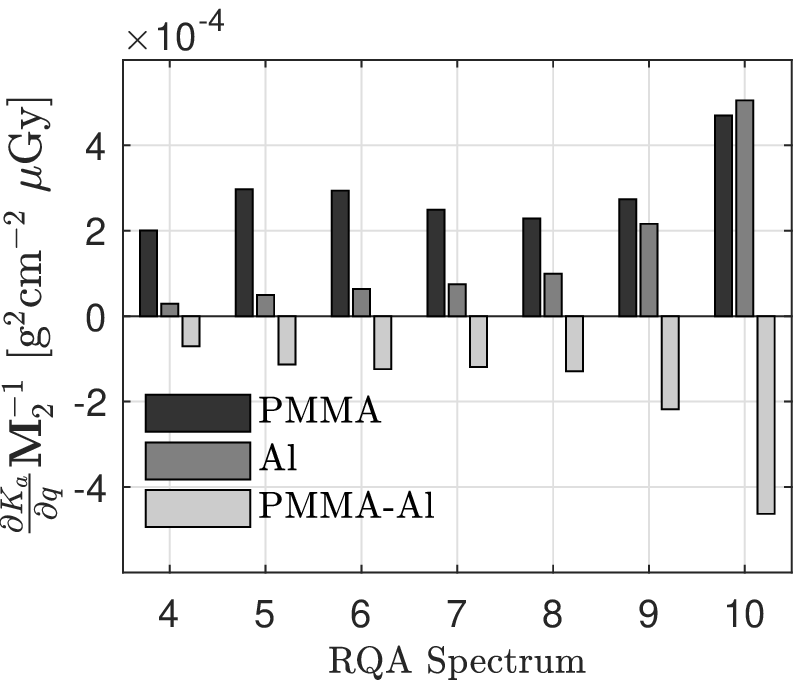}\includegraphics{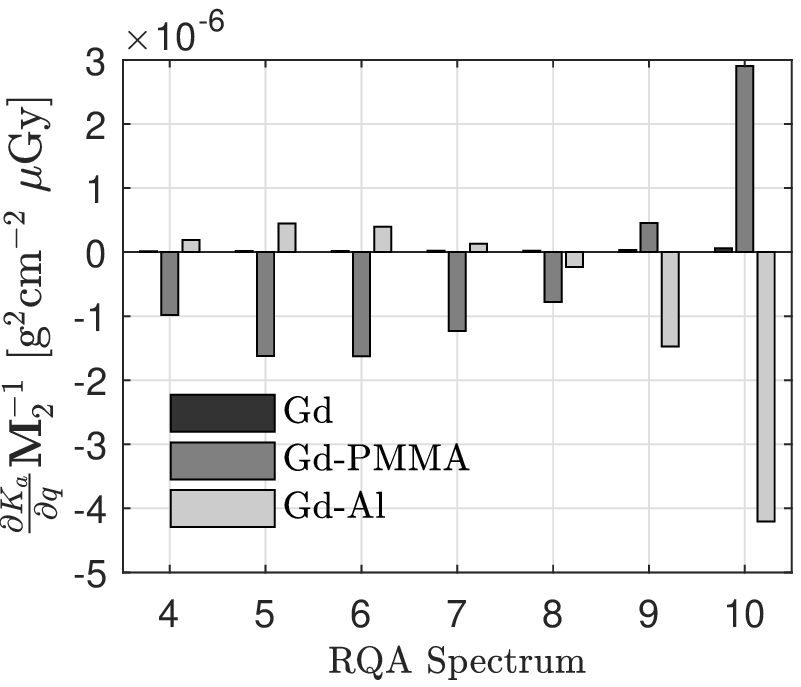}

\caption{\label{fig:Bar-plots-PMMA-Al-Gd}Bar plots of the noise power of basis-material
images for ideal SXDs for a three-material decomposition with PMMA,
Al and Gd basis materials. The left bar plot shows the NPS for PMMA
and Al, in addition to the cross NPS between PMMA and Al. The right
bar plot shows the NPS for Gd in addition to the cross NPS between
Gd and PMMA and between Gd and Al. Note that the y-scales of the two
bar plots are different.}
\end{figure*}
\begin{figure*}
\includegraphics{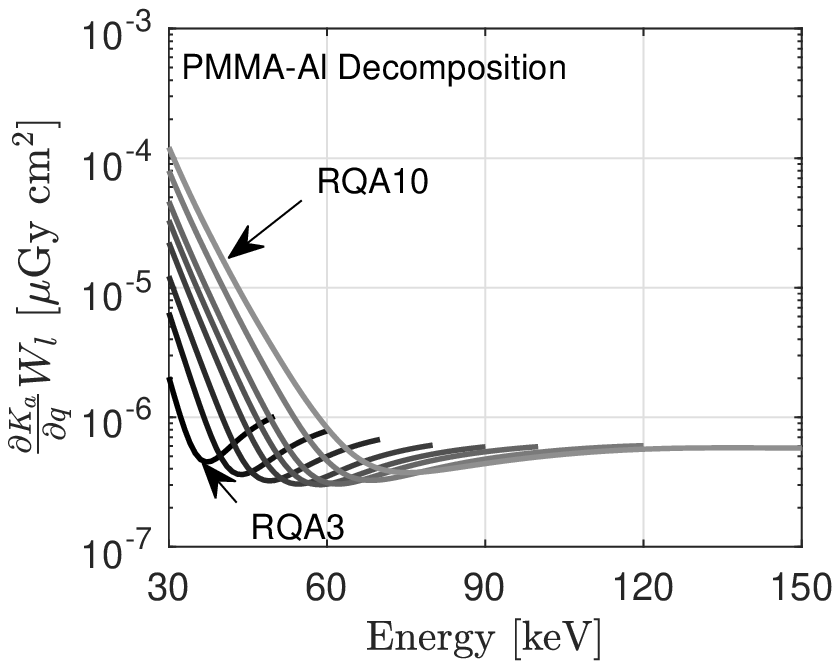}\includegraphics{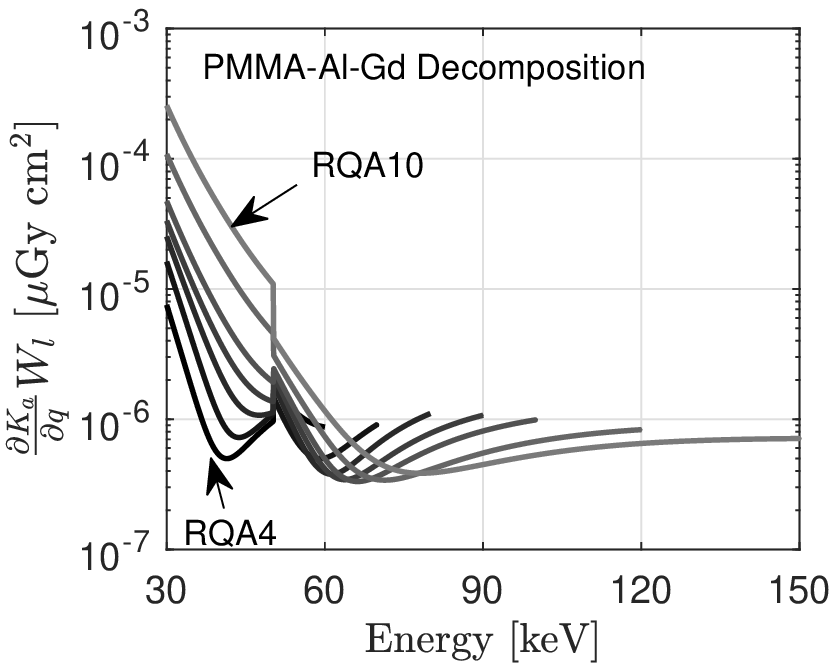}

\caption{\label{fig:Line_integral_SNR}Line-integral noise power normalized
by the air Kerma incident upon the detector. The left and right figures
correspond to two-material and three-material decompositions, respectively.
SNRs for the three-material decomposition are only shown for RQA spectra
with maximum energies greater than the K-edge energy of Gd.}
\end{figure*}
Example basis-material noise power spectra (i.e. the elements of $\mathbf{W}_{a}$)
are illustrated graphically in Fig.~\ref{fig:Bar-plots-PMMA-Al}
for the two normalizations described in Sec.~\ref{subsec:Calculations}.
Results are shown for a two-material decomposition with a basis set
consisting of PMMA and Al. The ``PMMA'' and ``Al'' bars represent
the noise power in the respective basis material images; the ``PMMA-Al''
bars represent the cross NPS between PMMA and Al basis materials.

The left bar plot of Fig.~\ref{fig:Bar-plots-PMMA-Al} shows the
inverse of the matrix of inner products ($\mathbf{M}_{2}^{-1}$) normalized
by the air Kerma incident on a 20-cm-thick volume of water ($K_{a,\mathrm{entrance}}$)
for tube voltages corresponding to the RQA series spectra in the right
figure. The left figure shows that, for a fixed entrance Kerma, basis
material noise power spectra and cross noise power spectra decrease
with increasing tube voltage. This result simply reflects the greater
attenuation of low-energy photons, which results in fewer quanta incident
on the detector for a fixed entrance Kerma.

The right bar plot shows $\mathbf{M}_{2}^{-1}$ normalized by the
air Kerma incident on the detector ($K_{a}$). The results in the
right figure are those which are required to normalize the DQE for
detection and quantification tasks, including pseudo-monoenergetic
imaging. For a fixed detector Kerma, the magnitude of basis-material
noise power spectra and cross noise power spectra increases with increasing
tube voltage. This is likely because the basis-material mass attenuation
coefficients are less distinct from each other at higher energies,
resulting in poorer conditioning of $\mathbf{M}_{2}$.

Figure~\ref{fig:Bar-plots-PMMA-Al-Gd} shows bar plots of basis-material
noise power spectra for a three-material decomposition with PMMA,
Al and Gd basis materials. Results are shown for x-ray spectra with
maximum energies greater than the K-edge energy of Gd. The noise power
spectra in Fig.~\ref{fig:Bar-plots-PMMA-Al-Gd} are normalized by
the air Kerma at the detector. The left bar plot of Fig.~\ref{fig:Bar-plots-PMMA-Al-Gd}
shows the NPS for PMMA and Al, in addition to the cross NPS between
PMMA and Al. The right bar plot shows the NPS of Gd, in addition to
the cross NPS between Gd and PMMA and between Gd and Al.

Comparing the left bar plot of Fig.~\ref{fig:Bar-plots-PMMA-Al-Gd}
with the right bar plot of Fig.~\ref{fig:Bar-plots-PMMA-Al} shows
that the addition of Gd to the basis set increases noise in the ideal
PMMA and Al basis images. This is expected because the inclusion of
a third basis material is an attempt to extract more information from
the energy-bin data. Comparing the left and right bar plots of Fig.~\ref{fig:Bar-plots-PMMA-Al-Gd}
shows that the noise in the ideal Gd basis image is much lower than
that in the PMMA and Al basis images. The Gd cross terms are also
much lower than the cross terms between PMMA and Al. These latter
two results are likely due to the presence of the Gd K-edge at $\sim$50-keV,
which results in a mass-attenuation coefficient with a substantially
different energy dependence than that of PMMA and Al.

\subsection{Pseudo-monoenergetic NPS ($W_{l,\text{ideal}}$)}

Examples of line-integral noise power spectra for ideal SXDs are illustrated
graphically in Fig.~\ref{fig:Line_integral_SNR}, in which noise
power spectra are normalized by the air Kerma incident upon the detector.
The x-axis represents the pseudo-energy at which the line integral
is evaluated. The left plot of Fig.~\ref{fig:Line_integral_SNR}
shows results for a two-material decomposition with PMMA and Al basis
materials; the right plot shows results for a three-material decomposition
with PMMA, Al and Gd basis materials.

For the two-material decomposition, the different curves (each corresponding
to a different spectrum) show similar trends, each having a minimum
between 30\,keV and the maximum energy of the x-ray spectrum. Although
not shown here, the minima occur at energies slightly less than the
average energies of the x-ray spectra. For example, the average energies
of the RQA7 and RQA9 spectra are 63\,keV ant 76\,keV, respectively,
and the corresponding minima occur at approximately 59\,keV and 70\,keV,
respectively. Similar results are observed for the three-material
decomposition, although the presence of the Gd K-edge introduces a
discontinuity in the NPS.

\section{Discussion}

We presented simple, closed-form expressions for the noise power spectra
of ideal SXDs for detection and quantification tasks. We showed that
a single matrix ($\mathbf{M}_{2}$) determines the NPS of ideal SXDs
for detection and quantification, including basis-material decomposition
and pseudo-monoenegetic imaging. The inverse of this matrix is proportional
(through the x-ray fluence) to the covariance matrix (per unit frequency)
of an unbiased GLS estimate of a perturbation ($\Delta\mathbf{a}$)
of basis material area densities. We showed that the elements of $\mathbf{M}_{2}$
are given by inner products of the mass-attenuation coefficients with
respect to the spectrum of photons incident on the x-ray detector.

The matrix $\mathbf{M}_{2}$ is determined uniquely by the x-ray spectrum
incident upon the detector and the basis set, and therefore can be
tabulated for standardized x-ray spectra and common basis sets, as
was done here. The resulting tables (in Appendix~A) can be used to
normalize experimental measurements of the frequency-dependent SNR
of SXDs for calculation of the task-dependent DQE for detection tasks
(Eq.~\ref{eq:DQE_D}) and quantification tasks, including basis-material
decomposition (Eq.~\ref{eq:DQE}) and pseudo-monoenergetic imaging
(Eq.~\ref{eq:DQE_E}). We did not tabulate values of the noise power
for pseudo-monoenergetic imaging, but these can easily be obtained
by combining the mass-attenuation coefficients of basis materials
with the tables in Appendix~A.

Whereas the DQE of a conventional, non-energy-resolving x-ray detector
is a single number, the DQE of an SXD is dependent on the imaging
task, so that it becomes necessary to distinguish between quantification
tasks (Eq.~(\ref{eq:DQE}) for basis images and Eq.~(\ref{eq:DQE_E})
for pseudo-monoenergetic images) or a detection task (Eq.~(\ref{eq:DQE_D})).
In a detection task, the objective is to determine whether a particular
feature is present in the image or not. In the context of spectral
x-ray imaging, this means aggregating the available information in
all energy bins or, after material decomposition, all basis material
images. Even if the signal difference to be detected is limited to
one single basis material, the ideal observer will still incorporate
data from the other basis material images, accounting for noise correlations
between these basis material images in an optimal way. In quantification
tasks, on the other hand, the objective is to measure a numerical
value as accurately and precisely as possible, e.g. the line integral
of one of the basis coefficients or of the attenuation at a particular
energy. Of course, it may then be of interest to assess the performance
for a detection task in such an estimated image, in which case Equation~(\ref{eq:DQE})
describes the detection performance relative to an ideal detector
when only a single basis image is available as opposed to Eq.~(\ref{eq:DQE_D})
which describes detection performance based on all basis images together.
Eq.~(\ref{eq:DQE}) can therefore be relevant for those detection
tasks where the other basis images can add confusion, such as when
using a K-edge image to detect elevated contrast agent concentrations
against a cluttered anatomical background.

Calculating the task-dependent DQEs in Eqs.~(\ref{eq:DQE}), (\ref{eq:DQE_E})
and (\ref{eq:DQE_D}) requires measurement of the NPS of individual
energy bins in addition to the cross NPS between energy bins, for
example using the approach described by Tanguay\emph{~et~al.}\citealp{Tanguay2021}
Also required for experimental DQE analysis is the matrix $\widehat{\mathbf{M}}$
with elements given by Eq.~(\ref{eq:Realistic_M}), which is proportional
to the MTF of energy bin $i$ for a spectrum that is weighted by the
mass-attenuation coefficient of basis material $b$. For basis sets
consisting of PMMA and Al, these elements could be measured using
the conventional slanted-edge approach\citealp{Samei1998} for spectra
transmitted through thin PMMA or Al absorbers. The development of
such an experimental framework is beyond the scope of this work, but
Persson~\emph{et~al.}\citealp{Persson2018,Persson2020} modeled
the DQE of cadmium telluride and silicon for quantification and detection
tasks for computed tomography imaging conditions. Their models predicted
that realistic CdTe SXDs produce DQEs $\sim$0.85 for water detection
tasks; for quantification tasks, the DQE drops to $\sim$0.35. Development
of an experimental framework for task-based DQE analysis is a focus
of ongoing research.

We chose to consider PMMA and Al basis materials because these materials
are widely available, relatively inexpensive, and convenient for experimentation.
While water and bone basis materials may be more clinically relevant,
water is cumbersome experimentally, and bone (or bone-equivalent plastic)
is not as readily available as Al. The tabulated matrix values can
also easily be transformed to another set of basis materials, if the
new set approximately lies within the span of the original ones. Specifically,
if we let $\left\{ \tfrac{\mu}{\rho}_{b^{\prime}}^{\prime}\left(E\right)\right\} _{b^{\prime}=1,\ldots,B}$
denote such a second set of basis materials, related to the original
basis materials by a transformation matrix $\mathbf{B}$ according
to $\tfrac{\mu}{\rho}_{b^{\prime}}^{\prime}\left(E\right)=\sum_{b}\tfrac{\mu}{\rho}_{b}\left(E\right)B_{bb^{\prime}}$,
$\mathbf{M}_{2}$ will be transformed into $\mathbf{M}_{2}^{\prime}=\mathbf{B}^{T}\mathbf{MB}$.
This formula can be used to calculate the quantification DQE for light
elements other than PMMA and Al, but if another K-edge element is
introduced, the elements of $\mathbf{M}_{2}$will have to be recomputed
using Eq.~(\ref{eq:correlation_matrix}).

We used RQA spectra for pragmatic reasons as well. Specifically, RQA
x-ray spectra are standardized and easily reproducible across different
laboratories. Secondly, RQA spectra have been adopted for assessment
of the DQE of conventional detectors. In practice, it will likely
be unnecessary to report a DQE for all the RQA spectra considered
in this work, and the lower-energy spectra may ultimately be irrelevant.
The most relevant RQA spectra for SXDs will likely be those produced
using tube voltages greater than 100\,kV, since higher tube voltages
tend to produce better image quality for SXDs.

\section{Conclusions}

The conclusions of this work are:
\begin{enumerate}
\item Normalization of the task-dependent DQE of spectroscopic x-ray detectors
requires knowledge of a matrix ($\mathbf{M}_{2}$) that is uniquely
determined by the x-ray spectrum incident upon the detector and the
mass-attenuation coefficients of the basis materials with respect
to which the signal to be detected or quantified is decomposed. (See
Eq.~\ref{eq:correlation_matrix}).
\item When divided by the air Kerma ($K_{a}$) incident upon the detector,
the diagonal elements of the matrix $(\partial K_{a}/\partial q)\mathbf{M}_{2}^{-1}$
{[}g$^{2}$cm$^{-2}$$\mu$Gy{]} (where $q$ {[}cm$^{-2}${]} represents
fluence) are equal to the noise power spectra of basis-material images
produced by an ideal spectroscopic x-ray detector; the off-diagonal
elements represent cross noise power spectra between basis materials.
\item Tables~\ref{tab:BM-PMMA-Al}, \ref{tab:BM-PMMA-Al-I} and \ref{tab:BM-PMMA-Al-Gd}
tabulate the elements of $(\partial K_{a}/\partial q)\mathbf{M}_{2}^{-1}$
for RQA x-ray spectra for selected basis sets, and can be incorporated
into an experimental framework for DQE analysis of spectroscopic x-ray
detectors, similar to the way tabulated values of the SNRs of RQA
x-ray fluences (in Tab.~\ref{tab:RQA}) are used in conventional
DQE analyses.
\end{enumerate}

\section{Conflict of Interest Statement}

M. Persson discloses past financial interests in Prismatic Sensors
AB (now part of GE Healthcare).
\begin{acknowledgments}
This work was supported by the Natural Sciences and Engineering Research
Council of Canada (NSERC) Discovery Grants program and MedTechLabs.
\end{acknowledgments}

\appendix

\section{Noise Power Tables}

Tables \ref{tab:BM-PMMA-Al}, \ref{tab:BM-PMMA-Al-I} and \ref{tab:BM-PMMA-Al-Gd}
list the basis-material noise power spectra and cross noise power
spectra for ideal SXDs for PMMA-Al, PMMA-Al-I and PMMA-Al-Gd basis
sets.
\begin{table}
\caption{\label{tab:BM-PMMA-Al}Noise power of an ideal SXD for PMMA-Al decomposition
tasks for RQA series x-ray spectra.}

\centering{}%
\begin{tabular}{cccc}
\hline
 & \multicolumn{3}{c}{$\partial K_{a}/\partial q\mathbf{M}_{2}^{-1}$ {[}g$^{2}$cm$^{-2}$$\mu$Gy{]}}\tabularnewline
Spectrum & PMMA & Al & PMMA-Al\tabularnewline
\hline
RQA3 & 1.33$\times$10$^{-4}$ & 1.64$\times$10$^{-5}$ & -4.54$\times$10$^{-5}$\tabularnewline
RQA4 & 1.32$\times$10$^{-4}$ & 2.65$\times$10$^{-5}$ & -5.75$\times$10$^{-5}$\tabularnewline
RQA5 & 1.34$\times$10$^{-4}$ & 3.70$\times$10$^{-5}$ & -6.84$\times$10$^{-5}$\tabularnewline
RQA6 & 1.53$\times$10$^{-4}$ & 5.52$\times$10$^{-5}$ & -8.97$\times$10$^{-5}$\tabularnewline
RQA7 & 1.75$\times$10$^{-4}$ & 7.40$\times$10$^{-5}$ & -1.11$\times$10$^{-4}$\tabularnewline
RQA8 & 2.02$\times$10$^{-4}$ & 9.67$\times$10$^{-5}$ & -1.37$\times$10$^{-4}$\tabularnewline
RQA9 & 2.68$\times$10$^{-4}$ & 1.53$\times$10$^{-4}$ & -1.98$\times$10$^{-4}$\tabularnewline
RQA10 & 3.32$\times$10$^{-4}$ & 2.16$\times$10$^{-4}$ & 2.63$\times$10$^{-4}$\tabularnewline
\hline
\end{tabular}
\end{table}
\begin{table*}[p]
\caption{\label{tab:BM-PMMA-Al-I}Noise power of an ideal SXD for PMMA-Al-I
decomposition tasks for RQA series x-ray spectra.}

\centering{}%
\begin{tabular}{ccccccc}
\hline
 & \multicolumn{6}{c}{$\partial K_{a}/\partial q\mathbf{M}_{2}^{-1}$ {[}g$^{2}$cm$^{-2}$$\mu$Gy{]}}\tabularnewline
Spectrum & PMMA & Al & I & PMMA-Al & PMMA-I & Al-I\tabularnewline
\hline
RQA3 & 3.34$\times10^{-4}$ & 2.04$\times10^{-5}$ & 8.18$\times10^{-9}$ & -7.14$\times10^{-5}$ & -1.18$\times10^{-6}$ & 1.80$\times10^{-7}$\tabularnewline
RQA4 & 1.48$\times10^{-4}$ & 3.23$\times10^{-5}$ & 1.46$\times10^{-8}$ & -4.78$\times10^{-5}$ & -4.83$\times10^{-7}$ & -2.91$\times10^{-7}$\tabularnewline
RQA5 & 1.43$\times10^{-4}$ & 1.06$\times10^{-4}$ & 3.90$\times10^{-8}$ & -9.31$\times10^{-5}$ & 5.85$\times10^{-7}$ & -1.64$\times10^{-6}$\tabularnewline
RQA6 & 3.47$\times10^{-4}$ & 4.69$\times10^{-4}$ & 1.43$\times10^{-7}$ & -3.73$\times10^{-4}$ & 5.25$\times10^{-6}$ & -7.67$\times10^{-6}$\tabularnewline
RQA7 & 8.98$\times10^{-4}$ & 1.32$\times10^{-3}$ & 3.63$\times10^{-7}$ & -1.06$\times10^{-3}$ & 1.62$\times10^{-5}$ & -2.13$\times10^{-5}$\tabularnewline
RQA8 & 2.31$\times10^{-3}$ & 3.38$\times10^{-3}$ & 8.66$\times10^{-7}$ & -2.77$\times10^{-3}$ & 4.28$\times10^{-5}$ & -5.33$\times10^{-5}$\tabularnewline
RQA9 & 1.23$\times10^{-2}$ & 1.73$\times10^{-2}$ & 3.98$\times10^{-6}$ & -1.46$\times10^{-2}$ & 2.19$\times10^{-4}$ & -2.61$\times10^{-4}$\tabularnewline
RQA10 & 3.60$\times10^{-2}$ & 4.86$\times10^{-2}$ & 1.05$\times10^{-5}$ & -4.18$\times10^{-2}$ & 6.11$\times10^{-4}$ & -7.12$\times10^{-4}$\tabularnewline
\hline
\end{tabular}
\end{table*}
\begin{table*}
\caption{\label{tab:BM-PMMA-Al-Gd}Noise power of an ideal SXD for PMMA-Al-Gd
decomposition tasks for RQA series x-ray spectra.}

\centering{}%
\begin{tabular}{ccccccc}
\hline
 & \multicolumn{6}{c}{$\partial K_{a}/\partial q\mathbf{M}_{2}^{-1}$ {[}g$^{2}$cm$^{-2}$$\mu$Gy{]}}\tabularnewline
Spectrum & PMMA & Al & Gd & PMMA-Al & PMMA-Gd & Al-Gd\tabularnewline
\hline
RQA3 & 1.67$\times10^{-3}$ & 6.64$\times10^{-3}$ & 2.83$\times10^{-5}$ & -3.23$\times10^{-3}$ & -2.09$\times10^{-4}$ & 4.33$\times10^{-4}$\tabularnewline
RQA4 & 2.00$\times10^{-4}$ & 2.90$\times10^{-5}$ & 1.40$\times10^{-8}$ & -7.06$\times10^{-5}$ & -9.81$\times10^{-7}$ & 1.88$\times10^{-7}$\tabularnewline
RQA5 & 2.97$\times10^{-4}$ & 4.93$\times10^{-5}$ & 1.61$\times10^{-8}$ & -1.13$\times10^{-4}$ & -1.62$\times10^{-6}$ & 4.45$\times10^{-7}$\tabularnewline
RQA6 & 2.94$\times10^{-4}$ & 6.36$\times10^{-5}$ & 1.88$\times10^{-8}$ & -1.24$\times10^{-4}$ & -1.62$\times10^{-6}$ & 3.97$\times10^{-7}$\tabularnewline
RQA7 & 2.49$\times10^{-4}$ & 7.48$\times10^{-5}$ & 2.03$\times10^{-8}$ & -1.19$\times10^{-4}$ & -1.23$\times10^{-6}$ & 1.30$\times10^{-7}$\tabularnewline
RQA8 & 2.29$\times10^{-4}$ & 9.91$\times10^{-5}$ & 2.28$\times10^{-8}$ & -1.29$\times10^{-4}$ & -7.77$\times10^{-7}$ & -2.32$\times10^{-7}$\tabularnewline
RQA9 & 2.74$\times10^{-4}$ & 2.16$\times10^{-4}$ & 3.42$\times10^{-8}$ & -2.18$\times10^{-4}$ & 4.54$\times10^{-7}$ & -1.47$\times10^{-6}$\tabularnewline
RQA10 & 4.70$\times10^{-4}$ & 5.05$\times10^{-4}$ & 6.12$\times10^{-8}$ & -4.63$\times10^{-4}$ & 2.91$\times10^{-6}$ & -4.20$\times10^{-6}$\tabularnewline
\hline
\end{tabular}
\end{table*}

\section{\label{sec:Table-of-Symbols}Table of Symbols and Relation to Previous
Work}

Table~\ref{tab:Symbols} lists some of the important quantities used
in this work together with the corresponding quantities in Persson~\emph{et~al.}\citealp{Persson2018}
\begin{table*}
\caption{\textcolor{blue}{\label{tab:Symbols}}Summary of important quantities
together with the corresponding quantities in Persson~\emph{et~al.}\citealp{Persson2018}.
In the right column, \textrm{$\rho_{b}$} denotes the density of basis
material \textrm{$b$ }and \textrm{$d_{i}$} denotes the position-independent
background counts in bin $i$. Subscript indices are consistent with
the present work rather than Persson~\emph{et~al.}\citealp{Persson2018}\textrm{.}}
\begin{tabular}{cccc}
\hline
Description & Symbol & Unit & Corresponding notation in \citealp{Persson2018} \tabularnewline
\hline
Signal difference in frequency domain & $\widehat{\Delta a}_{b}$ & g & $\rho_{b}\Delta\tilde{A}_{b}$\tabularnewline
Normalized bin counts difference in frequency domain & $\widehat{\Delta L}_{i}$ & cm$^{2}$ & $-\Delta D_{i}^{s}\Delta_{x}\Delta_{y}/d_{i}^{s}$\tabularnewline
Inner product matrix element & $\left[\mathbf{M}_{2}\right]_{b,b^{\prime}}$ & cm$^{4}$g$^{-2}$ & N/A\tabularnewline
Transformation matrix element & $\widehat{M}_{i,b}$ & cm$^{2}$g$^{-1}$ & $-\frac{1}{d_{i}^{s}}\frac{\Delta_{x}\Delta_{y}}{\rho_{b}\overline{q}^{\mathrm{tot}}}\frac{\partial\overline{q}^{\mathrm{tot}}}{\partial A_{b}}H_{i,b}^{\mathcal{B}}$\tabularnewline
Incident fluence per unit energy & $\phi$ & keV$^{-1}$cm$^{-2}$ & $\overline{q}$\tabularnewline
Incident spectrum difference in frequency domain & $\widehat{\Delta\Phi}$ & keV$^{-1}$ & $-\Delta\overline{Q}$\tabularnewline
Energy bin counts difference in frequency domain & $\widehat{\Delta c}_{i}$ & cm$^{2}$ & $-\Delta\overline{D}_{i}^{s}\Delta_{x}\Delta_{y}$\tabularnewline
Transfer function of energy bin & $T_{i}$ & cm$^{2}$ & $\Delta_{x}\Delta_{y}H_{i}$\tabularnewline
Basis material noise power matrix element & $\left[\mathbf{W}_{\mathbf{a}}\right]_{b,b^{\prime}}$ & g$^{2}$/cm$^{2}$ & $\overline{q}^{\mathrm{tot}}{}^{2}\frac{\left[\left.\mathsf{NEQ}^{\mathcal{B}}\right.^{-1}\right]_{b,b^{\prime}}}{\frac{\partial\overline{q}^{\mathrm{tot}}}{\partial A_{b}}\frac{\partial\overline{q}^{\mathrm{tot}}}{\partial A_{b^{\prime}}}}\rho_{b}\rho_{b^{\prime}}$\tabularnewline
\hline
\end{tabular}
\end{table*}

\end{document}